\documentstyle[12pt,epsf]{article}
\newcommand{\bea}{\begin{eqnarray}}
\newcommand{\eea}{\end{eqnarray}}
\newcommand{\be}{\begin{equation}}
\newcommand{\ee}{\end{equation}}
\newcommand{\bi}{\bibitem}
\newcommand{\D}{\mbox{$\Delta$}}
\newcommand{\mitD}{\mbox{${\mit \Delta}$}}
\renewcommand{\d}{\mbox{$\delta$}}
\renewcommand{\t}{\mbox{$\theta$}}
\renewcommand{\o}{\mbox{$\omega$}}
\renewcommand{\a}{\mbox{$\alpha$}}
\newcommand{\s}{\mbox{$\sigma$}}

\newcommand{\ep}{\mbox{$\epsilon$}}
\newcommand{\p}{\mbox{$\psi$}}
\newcommand{\bfk}{\mbox{${\bf k}$}}
\newcommand{\bfR}{\mbox{${{\bf R}}$}}

\newcommand{\naR}{\mbox{$\nabla_{\bf R}$}}
\newcommand{\xipa}{\mbox{$\xi_{\parallel}$}}
\newcommand{\xipe}{\mbox{$\xi_{\perp}$}}
\newcommand{\xia}{\mbox{$\xi_{ani}$}}

\begin{document}
\begin{center}
{\bf \large Thermal Fluctuations in d-wave Layered Superconductors}\\
{\today}
\vspace{.5 cm}

{ Muriel Ney-Nifle\footnote{e-mail : ney@solrt.lps.u-psud.fr,\\
permanent address : Laboratoire de Physique,
 Ecole Normale Sup\'erieure,
46 all\'ees d'Italie, 69364 Lyon cedex 07, France}
and Marc Gabay}
\vspace{.5 cm}

Laboratoire de Physique des Solides
\footnote{Laboratoire associ\'e au CNRS}, Universit\'e de Paris-Sud,\\
B\^atiment 510, 91405 Orsay Cedex, France
\vspace{2 cm}

{\bf Abstract}
\end{center}

We study the thermal fluctuations of anisotropic order parameters (OP) in
layered superconductors. In particular, for
copper oxides and a d-wave OP, we present some
experimental consequences of fluctuations in the direction normal
to the layers. It is
shown that the c-axis penetration depth $\lambda_c$ can have a
"disorder-like" quadratic temperature dependence at low temperature.

The fluctuations are analyzed in the framework of a Lawrence-Doniach model
with an isotropic Fermi surface. Anisotropies pin
the orientation of the OP to the crystallographic axes of the lattice. Then
we study an extended t-J model that fits Fermi suface data of bilayers
$YBCO$ and $BSCCO$. This leads to a d-wave OP with two possible orientations
and, including the thermal fluctuations, yields the announced
temperature dependence of $\lambda_c$.
Furthermore a reservoir layer is introduced. It implies a finite
density of states at the Fermi energy which is successfully compared
to conductance and specific heat measurements.

\vspace{1.2 cm}
\vspace{3 cm}

\section{Introduction}

Despite an enormous amount of theoretical and experimental
research devoted to High Temperature Superconductors (HTSC) in the past
decade, many important issues still remain unsettled. In particular there is
as yet no explanation for the pairing mechanism at work in these materials.
Early on Anderson suggested \cite{A} that spin fluctuations might
play a critical role for the superconductivity of copper oxides. These
fluctuations have been analyzed for instance in the framework of the t-J model
\cite{rice} and a natural
candidate to represent the superconducting state in that model is a d-wave
order parameter (OP).

Many low temperature experiments seem to suggest that a
d-wave symmetry is a strong possibility : the variation of the penetration
depth with $T$ \cite{hardy,mao} and $H$ \cite{maeda}, Knight Shift
and $1/T_1T$ data \cite{taki}, angle resolved photoemission spectra
\cite{shen}, the variation of
$C_V$ with $T$ and $H$ \cite{moler}. Yet these experiments do not probe
directly the OP so that one cannot rule out other symmetries or
mixtures of symmetries \cite{V}.
 Precise measurements of the single particle density of
state (DOS) near the Fermi energy in the superconducting
state ought to give some
clue about the pairing symmetry. Tunneling experiments however have been
fairly inconclusive thus far, leading to s-wave, d-wave and, in the case
of $YBCO$, to the conductance of a system with finite DOS at the
Fermi surface \cite{L,dynes}.
Materials issues have made it difficult to
sort out genuine features from experimental artefacts. The observation of spin
fluctuations well below $T_c$ seen by inelastic neutron scattering and a
linear in $T$ dependence of $C_V$ once the phonon contribution is subtracted
\cite{aeppli} have been interpretated as evidence for gapless
superconductivity.

Despite the
fact that a clear-cut answer still remains elusive for high $T_c$ materials,
an interesting by-product of these studies has been a renewed interest in
coherence effects in d-wave superconductors. Sigrist and Rice indeed suggested
that d-wave superconducting currents would display a strong spatial anisotropy
at a boundary \cite{SR}.
 This anisotropy should show up in Josephson experiment setups
\cite{W}.
 Ambiguous results have been obtained, although
Tsuei-Kirtley's experiment strongly supports a scenario
where one has nodes for the OP (consistent with either d-wave
or extended s-wave).
Furthermore, the layer structure caracterizing $CuO_2$
compounds should affect tunneling and Josephson currents as it does  for
vortex (thermo)-dynamics below $T_c$ and for superconducting fluctuations
above $T_c$, where a Lawrence-Doniach model is used \cite{LD}.
 These reasons led us to
investigate the issues of layering and of anisotropies in d-wave
superconductors. Our results show that these effects may play a significant
role in the interpretation of  tunneling and
penetration depth types of experiments.

Our paper is organized as follows : in section 2.1 we present the
Landau-Ginzburg free energy of a single layer of a d-wave superconductor with
an isotropic Fermi surface by solving the Gorkov equations.
We calculate the energy cost of the fluctuations in
{\it the orientation of the nodal lines}, denoted $\a$, of an OP
 with the $d_{x^2-y^2}$ symmetry. These fluctuations can be
seen as admixing $d_{x^2-y^2}$ and $d_{xy}$ terms
and appear in gradient terms of the free energy.
We also show that the d-wave symmetry results in
anisotropies in gradients of $\a$ and of the phase of the OP. These do not
modify the critical behaviour near $T_c$ but have an influence in the London
regime at low $T$. Then, introducing anisotropies in the Fermi surface,
we see that this tends to orient the nodal lines at $45^{\circ}$
with respect to crystallographic axes. We next consider the case of weakly
coupled d-wave superconducting sheets and we
derive the corresponding Lawrence-Doniach free energy (see section 2.2).
The energy cost of the fluctuations of $\a$ is once again
computed and gives, in particular, the Josephson current
perpendicular to the layers. We find that the fluctuations in
the c-direction are limited to a coherence length.

In section 3 we consider
 a t-J model that reproduces a reasonable  band structure
 for the bilayer compounds $YBCO$ and also $BSCCO$.
It is found that the nodal lines can now take on two positions
\cite{KL}. This seems
in agreement with recent experiments on $BSCCO$ \cite{ding}.
We add a hopping energy $t_n$ that couples two bilayers
via a charge reservoir; depending on the strength of $t_n$ one may go from
a more 3D structure pertaining to $YBCO$ to a more 2D structure relevant
for $BSCCO$.
We show how these modify the DOS which is determined analytically
near the Fermi surface.
It starts linearily, as in case with pure $d_{x^2-y^2}$
OP, with an offset that vanishes with $t_n$.
This offset is observed in specific heat measurements \cite{KAM}
that are in quantitative agreement with our result
and also in conductance measurements \cite{L,dynes}.

The existence of several
orientations of the nodal lines enhances the effect
of the thermal fluctuations pointed out above:
they give a mechanism that permits the OP to change its
orientation from one layer to another.
This has experimental implications that we discuss in section 3.3, in
particular in the case of c-direction
penetration depth measurements. Fluctuations
in the orientation of the OP in the c-direction
can change the T dependence of $\lambda_c$ from a
linear to a quadratic behaviour at low T, even
in pure samples where it is linear in the a- and b-directions.
This is consistent with experiments on $YBCO$ \cite{lambda1,guerin}.

Section 4 is a brief conclusion.

\section{Fluctuations of d-wave order parameters : phenomenology}

The model under consideration is the one introduced by
Lawrence and Doniach (LD) \cite{LD} for weakly coupled layered
 superconductors. Here we shall first study the motion of
electrons within a layer and then consider the Josephson
tunnel that couples adjacent layers.

We introduce the ``nodal angle'' which define the orientation
of the nodes of the  $d_{x^2-y^2}$ OP within a layer. It fluctuates both in
the direction parallel and perpendicular to the layers and we determine the
length scales of these fluctuations.

\subsection{Free energy density in a layer}

We consider a d-wave OP $\mitD_n (\bfk , \bfR )$ where
$\bfR$ is the center-of-mass coordinate
of a Cooper pair, $n$ the layer's number  and $\bfk$ the relative
coordinate in the Fourier space, $\bfk=(k_x,k_y)=
(|\bfk| \cos \t, |\bfk| \sin \t)$. We choose for the
$\bfk$-dependence of $\mitD$ the form ${\hat k}_x^2 - {\hat k}_y^2
=\cos (2\t)$ which has nodal lines, $k_x=\pm k_y$. If $\a_n(\bfR)$
denotes the relative angle of these lines with respect to
fixed axes one has
\be\label{D1}
\mitD_n(\bfk,\bfR ) = \D_0(T) \cos[2(\t - \a_n(\bfR ))]
 {\rm e}^{i\phi_n(\bfR )}
\ee

This can be expanded as
\be\label{D2}
\mitD_n(\bfk,\bfR ) = \D^n_1(\bfR ) \p_1(\t) +
\D^n_2(\bfR ) \p_2(\t)
\ee
where the $\D^n_i$ are two related gap functions
\be\label{delta}
\left\{
\begin{array}{ll}
\D^n_1(\bfR )=\D_0 \;\cos(2 \a) \;{\rm e}^{i\phi}\\
\D^n_2(\bfR )=\D_0 \;\sin(2 \a) \;{\rm e}^{i\phi}
\end{array}
\right.
\ee
and the $\p_i^n$ have the form of the d-wave pairing called
$d_{x^2-y^2}$ and $d_{xy}$
\be\label{psi}
\left\{
\begin{array}{ll}
\p ^n_1(\t)=\cos(2 \t) = {\hat k}_x^2 - {\hat k}_y^2\\
\p ^n_2(\t)=\sin(2 \t) = 2 {\hat k}_x {\hat k}_y
\end{array}
\right.
\ee
This is compatible with the factorized pairing potential
$V(\bfk-\bfk') = V \sum \p_i(\bfk) \p_i(\bfk')$.
The form (\ref{D2}) for the OP can  be generalised to other
symmetries or mixture of symmetries. A relevant example
is the case with a $s+d$ OP (e.g. when one has orthorhombicity)
which is given by (\ref{D2}) plus
a term $\D_3\p_3$ having the desired s-wave symmetry.
We mention later the resulting corrections to the free energy.

Then we shall write the GL equations for $\mitD_n$ where
one integrates over $\bfk$. We follow a method
analogous to the one used previously for mixed OP
 \cite{R} except that we introduce the nodal
angle $\a_n(\bfR)$ and we consider a different
set of $\{ \p_i, \D_i \}$. The derivation of these
equations may be done in two
steps. The first concerns any OP and generalizes the
s-wave pairing equations. Since this has been shown
several times (see e.g. \cite{G}) we only sketch the
calculation in Appendix A. The second step specializes
to the d-wave pairing considered above (\ref{D1}). The whole
procedure gives two coupled  gap equations for the $\D_i^n$
in which, in the limit $T\rightarrow T_c$, we retain only
the following terms
$\D, \nabla^2\D, \nabla^4\D$ and $\D^3$. We get
\bea\label{eqgap}
\D_1^n& =& a<\p_1^2>\D_1^n -
{b \over T_c^2}<\p_1^4>|\D_1^n|^2\D_1^n -
3{b \over T_c^2}<\p_1^2\p_2^2>|\D_2^n|^2\D_1^n
\nonumber\\
      &  &+\eta\; v_F^2\; ( <\p_1^2{\hat k}_x^2>\nabla^2_x +
   <\p_1^2{\hat k}_y^2>\nabla^2_y ) \;\D_1^n
\nonumber\\
&&+c\;v_F^4 \;( <\p_1^2{\hat k}_x^4>\nabla^4_x +
   <\p_1^2{\hat k}_y^4>\nabla^4_y +
   6<\p_1^2{\hat k}_x^2{\hat k}_y^2>\nabla^2_x \nabla^2_y ) \;\D_1^n
\nonumber\\
&&+c\;v_F^4\; ( 4<\p_1\p_2{\hat k}_x^3{\hat k}_y>\nabla^3_x \nabla_y +
   4<\p_1\p_2{\hat k}_x{\hat k}_y^3>\nabla_x \nabla^3_y ) \;\D_2^n
\eea
In (\ref{eqgap}), $<...>$ denotes
an integration over $\t$ and the various letters $a, b$, ..., denote the
integration and summation $\int d|\bfk | \sum_{\o} ...$. The Fermi
surface being circular
the energy only depends on $|\bfk |$ and the integration
on $\t$ affects only the $\p_i$ and the Fermi velocity
[i.e. $(k_x,k_y)=mv_F(\cos\t,\sin\t)$].
One gets a similar equation, upon making the substitution
$\D_1 \leftrightarrow \D_2$ (and $\p_1\leftrightarrow \p_2$),
 which has the same coefficients
$a, b, \eta$ and $c$. In these equations the d-wave pairing
manifests itself in the $<\p_i^j>$-dependence of the
coefficients (while a non zero value of $\a_n$
introduced two gaps $\D_i^n$).

It appears that the contribution
of the fourth order in $\nabla$ is anisotropic
(see the c-terms in (\ref{eqgap})). In the case
of s-wave pairing these terms can be  combined
to give a single isotropic contribution to the free energy,
as it should.

Finally the GL free energy density of the $n$th layer is given by
\be\label{fn}
f_n = f_0 + f_{ani}
\ee
with
\begin{eqnarray*}
f_0& =& -A(1-{T \over T_c}) \;( |\D_1^n|^2+|\D_2^n|^2 ) +
{B \over {2 T_c^2}} \;\;( |\D_1^n|^2+|\D_2^n|^2 ) ^2
\\
   &  &+E\;({hv_F\over T_c})^2 \;( |\naR \D_1|^2 + |\naR \D_2|^2 )
\end{eqnarray*}
and the anisotropic part
\begin{eqnarray*}
f_{ani}&= &\sum_{i=1,2} ({hv_F\over T_c})^4\;[
\;C_i\;( |\nabla_x^2\D_i^n|^2 +
|\nabla_y^2\D_i^n|^2) +
   \;C'_i \; |\nabla_x\nabla_y\D_i^n|^2 \;]
\\
   &  &+{C \over 2}\; ({hv_F\over T_c})^4 \;|
		\;\nabla_x\D_1^n\nabla_y^3\D_2^{n*} +
		\nabla_x\D_1^{n*}\nabla_y^3\D_2^{n} -
		\nabla_x^3\D_1^n\nabla_y\D_2^{n*} -
		\nabla_x^3\D_1^{n*}\nabla_y\D_2^{n}
\\
&& \;\;\;\;\;\;\;\;\;+(\D_1 \leftrightarrow \D_2)\;|
\end{eqnarray*}
where we used (\ref{psi}) and the circular model
to calculate the integrations over $\t$ which we then
incorporated in the coefficients $A, B, ...$.

This free energy density has been obtained
keeping only the leading term in the small-$|\bfk\ |$
expansion which yields the from (\ref{D1}) for the OP and
an isotropic Fermi surface (FS). Within this restriction
we got an expression that does not depend on $\a_n({\bf R})$
at the mean-field level. In other words,
the free energy of one layer is minimised by any uniform
$\a_n$.

 Let us now consider the anisotropy effects.
They are two sources of anisotropy, one in
the OP, the other in the Fermi energy.
The effect of the first one is illustrated by the presence
of $f_{ani}$ which implies that the fluctuations of
 $\a_n({\bf R})$ (and also of $\phi_n({\bf R})$)
are anisotropic in a layer. The fact that this
anisotropy appears to order $\naR^4\D^2$ is due to the
particular form of the OP (another form
yields another order, e.g. an additional $s$-wave OP
gives anisotropic terms to order $\naR^2\D^2$ \cite{V}).

One can verify that the anisotropic terms
$f_{ani}$ contain gradients of $\a$ and of $\phi$ which are
coupled (this coupling being
absent at lower orders). These terms are negligible
in the high temperature region (near $T_c$) but
come into play at lower $T$. In the London regime,
it could be interesting to study the implication of
this coupling on the vortex state. As one
replaces gradients by $\nabla - {ei \over h} {\bf A}$
the term to the quadratic order takes on the form
\be
\D_0^2E\{ ({hv_F\over T_c})^2 [4(\naR\a_n)^2
   +({\bf A}+\naR \phi_n)^2]
\ee
In addition to the kinetic energy of the current
one gets a rigidity term in the orientation of
the OP, $\a_n$. Thus the fluctuations
of $\a_n$ are relevant as long as $\a_n$ is not
fixed relatively to the crystallographic axes.

Indeed the introduction of an anisotropy in the FS
 due to a crystal structure can pin in the orientation $\a_n$
 and also generate anisotropic fluctuations
of $\a_n$. To show this on an example,
we consider an elliptic FS, i.e. $k_x^2+k_y^2+\epsilon k_y^2$.
To first order in $\epsilon$ the anisotropic terms
now come in to order $\nabla^2 \D^2$ while to
order $\epsilon^2$ the terms $|\D_1^n|^2$ and $|\D_2^n|^2$
have different coefficients. This means that
the term to order $|\D|^2$ now depends on $\a_n$ which will
be fixed by the minimisation of the free energy.
In this case there are still anisotropic terms in the
Free energy (in gradients of the phase of the OP)
which are relevant in the London regime.

\subsection{Free energy density for layered superconductors}

Following \cite{LD} we now add a transverse component
to the electron energy of the normal state
\be\label{eps}
\epsilon (\bfk) = {{k_x^2+k_y^2} \over {2 m}} + K\; \cos(k_zd)
\ee
where $d$ is the interlayer spacing. We performed an
expansion of the gap equations in the limits
$T-T_c<<1$ and $K<<T_c$ (see Appendix A).
The free energy has the general form
$\D F=\sum_n \int d\bfR (f_n+f_{n,n+i})$ with $f_n$ given
by (\ref{fn}) and
\bea\label{fn1}
f_{n,n+i}&=&E\;{K^2\over T_c^2}\;( |\D_1^n-\D_1^{n+1}|^2 +
    |\D_2^n-\D_2^{n+1}|^2)
\nonumber\\
    &&+12\;C\;{K^2\over T_c^2}\;({hv_F\over T_c})^2\;(
       |\naR\;(\D_1^n-\D_1^{n+1})|^2+|\naR\;(\D_2^n-\D_2^{n+1})|^2\;)
\nonumber\\
    &&+{3\over 32}\;C\;{K^4\over T_c^4}\;\sum_{i=1,2}\;
      ( \;13|\D_i^n-\D_i^{n+1}|^2
\nonumber\\
    &&\;\;\;\;\;\;-\;8|\D_i^n-\D_i^{n+2}|^2
      +\;3|\D_i^n-\D_i^{n+3}|^2-\;{1\over 2}|\D_i^n-\D_i^{n+4}|^2\;)
\eea
Thus in this LD model thermal fluctuations
give rise to variations of the value of the nodal angle $\a_n$.
One sees in (\ref{fn}) and (\ref{fn1}) that $\a_n$ fluctuates,
both within a layer and from one layer to another, around
a constant but arbitrary value (in $f_0$
the first two terms do not depend on $\a_n$). The energy
cost of these fluctuations is given by
\be\label{df}
\d f=\D_0^2\;E\;\{ 4\;({hv_F\over T_c})^2 \;(\naR\a_n)^2
   \;+\;2\;{K^2\over T_c^2}\;( 1
-\cos[2(\a_n-\a_{n+1})]\cos(\phi_n-\phi_{n+1})) \}
\ee
which was determined from the gradient terms of $f_n$ and $f_{n,n+i}$
with the simplifying assumption that $\naR \phi _n= 0$. We wrote only
the dominant terms, that is, the first line in (\ref{fn1})  and the
 part $f_0$ of $f_n$ but one must keep in mind the presence
 of anisotropic terms. from (\ref{df}) one gets a current density
between two adjacent layers
\be
j_{n,n+1} \sim \cos [2(\a_n-\a_{n+1})]\sin(\phi_n-\phi_{n+1})
\ee
Although this is a current that flows perpendicularly to
the layers, it has the same behaviour with $\a_n-\a_{n+1}$
as the current across a tunnel junction in a layer \cite{bruder}.
Note that in a model where $\a_n - \a_{n+1}$ can become arbitrarily
large (i.e. when $\a_n$ is not pinned by anisotropies
in the FS), the fluctuations can supress the tunneling in the
c-direction ($j_{n,n+1}$ vanishes on average).

Now to study
the energetic balance of the fluctuations of $\a_n$
 one has to compare the two characteristic lengths,
$\xipa$ and $\xipe$, that
 come from $\d f$ and a third one, $\xia$, from $f_{ani}$ :
\be\label{xiper}
{\xipe^2 \over \xipa^2} =
{{K^2\;d^2} \over {2\;(hv_F)^2}}
\ee
and
\be\label{xiani}
{\xia^2 \over \xipa^2} =
{AC\over E^2} ({hv_F\over T_c})^2 (1-{T\over T_c})
\ee
where $AC/E^2$ is a numerical factor.
The LD model is valid within a regime of temperature given by
the limits
\be\label{limit}
1-{T\over T_c}\;<<\;1\;\;\;{\rm and}
\;\;\xipe\;\sim\;(1-{T\over T_c})^{-1/2}\;<<\;d
\ee
There, the intra-layer length $\xia$ is of order of $\xipa$,
while the inter-layer length, $\xipe$, is much smaller
yielding the fluctuations of the OP in the c-direction inexpensive.
We study some experimental implications of the fluctuations
in the c-direction in section 3.

The above analysis does not fix the value of $\a_n$
in the absence of anisotropies of the FS linked
to the crystallographic structure of these materials. To remedy this
 one has to consider a microscopic model that specifies a
reasonable band
structure (a step in that direction was taken in section 2.1).
We choose to start from the t-J model since it is
a natural candidate for a d-wave
superconducting phase \cite{rice}.

\section{The t-J model for bilayers coupled
via a reservoir}

In the usual t-J model one considers the pairing
of nearest-neighbor fermions on a square lattice
with an exchange energy $J_{\parallel}$ and a quasiparticle hopping
energy $t_{\parallel}$.
This leads \cite{rice,UL} to a normal state with
a Fermi surface (FS) of square shape at
half filling and favors a variational d-wave superconducting state
(in fact the lowest energy state corresponds to an $s+id$ OP at half
filling and to a d-wave OP away from half filling)
 There the OP has the $d_{x^2-y^2}$ symmetry
of (\ref{D1})  with a nodal angle $\a=0$,
( nodes lie at $45^{\circ}$ with respect to the crystallographic axes ).
Recently it was shown \cite{KL} that additional terms were needed
to fit experimental data on $BSCCO$:
Kuboki and Lee used a bilayer structure
with an exchange energy between two layers, $J_{\perp}$, and
a hopping term, $t_{\perp}$. In agreement with ARPES experiments \cite{bscco}
one gets a FS with two bands \cite{KL}.
Furthermore the OP has two nodal angles at $\a=\pm \a_0$
which may have been observed in $BSCCO$ \cite{ding} ($\a_0\sim 10^{\circ}$).
Here we shall go further by adding terms that describe the
FS and the density of states (DOS) of $YBCO$.
We include a pairing along the
diagonals in a layer, $J_d$, and a hopping term, $t_n$, that couples two
bilayers via a charge reservoir, see figure 1.

\subsection{Fermi surface and gap}

The slave-boson mean-field Hamiltonian of the t-J model
describes electrons
with forbidden doubled occupancy on a site :
the creation of an electron on site $i$ in the layer $s$
consists of the anihilation of a hole operator
having the charge of the electron $b_i^s$
and the creation of a fermion operator having the spin $\s$
of the electron $f_{i,\s}^{s+}$
($s=1,2$ for the bilayer and $s=3$
for the reservoir). To simplify we
shall set $b_i^{s+}b_j^s=\d$, i.e. the number of
empty sites gives the doping rate.

The hamiltonian is given by
\be\label{ham}
{\cal H}\;=\;-\sum_{r,s}\sum_{i,j}[ \;\d\;t_{ij}\;f^{r+}_{i\s}f^s_{j\s}
+\;J_{ij}
\;(\vec {S}^r_i.\vec {S}^s_j\;-\;{1\over 4}n_in_j) \;]
\ee
with the constraint
$\sum f^+_{i\s}f_{i\s} =1-\d$ and $\vec {S}^r_i=f^{r+}_{i\a} \vec {\s}
_{\a\beta} f^r_{j\beta}$, the
$t_{ij}$ and $J_{ij}$ are shown in figure 1.
Within mean-field approximation we introduce various OP, in particular we
restrict our analysis to a d-wave coupling in the layers and to an s-wave
coupling between layers :
\bea
J_{\parallel}\sum_{\s} \;\s< f^{r+}_{i\s}f^{r+}_{j-\s}>=+(-)\D_{\parallel}
&{\rm with}&j=i \pm x(\pm y)\;\;{\rm and}\;\;r=1,2
\nonumber\\
J_{d}\sum_{\s} \;\s< f^{r+}_{i\s}f^{r+}_{j-\s}>=+(-)\D_{d}
&{\rm with}&j=i \pm x \pm y(\mp x \mp y)\;\;{\rm and}\;\;r=1,2
\nonumber\\
J_{\perp}\sum_{\s} \;\s< f^{1+}_{i\s}f^{2+}_{j-\s}>=\D_{\perp}
&{\rm with}&j=i+ z
\nonumber\\
\eea
in the particle-particle channel and
\bea
\sum_{\s} \;< f^{r+}_{i\s}f^r_{j\s}>=\chi_{\parallel}
&{\rm with}&j=i \pm x\;{\rm or}\;\pm y
\nonumber\\
\sum_{\s} \;< f^{r+}_{i\s}f^r_{j\s}>=\chi_{d}
&{\rm with}&j=i \pm( x \pm y)
\nonumber\\
\sum_{\s} \;< f^{1+}_{i\s}f^2_{j\s}>=\chi_{\perp}
&{\rm with}&j=i+z
\nonumber\\
\sum_{\s} \;< f^{r+}_{i\s}f^3_{j\s}>=\chi_{n}
&{\rm with}&j=i+z
\nonumber\\
\eea
in the particle-hole channel ( here too $r=1,2$ ).

As usual \cite{rice,UL} we write the Hamiltonian terms
depending on the OP in momentum space
\bea\label{hk}
{\cal H}_{\D}=\sum _{\bf k} F^+({\bf k}) . M({\bf k}) . F({\bf k})+
{3 N\over 8} (2{\D^2_{\parallel}\over J_{\parallel}}
+2{\D^2_d\over J_d}+{\D^2_{\perp}\over J_{\perp}})
\\
{\rm with}\;\; F({\bf k})=(f^1_{{\bf k},\uparrow},f^1_{-{\bf k},\downarrow},
f^2_{{\bf k},\uparrow},f^2_{-{\bf k},\downarrow},
f^3_{{\bf k},\uparrow},f^3_{-{\bf k},\downarrow}),
\nonumber\\
M=
\left(
\begin{array}{rrrrrr}
\xi_1 	& \D_1 	& {\tilde t}_{\perp}	& \D_{12} 	& {\tilde t}_n 	& 0 \\
 	& -\xi_1 &\D_{12} & -{\tilde t}_{\perp} & 0 	&  -{\tilde t}_n \\
 	&  	& \xi_1 	& \D_1 		& {\tilde t}_n & 0 \\
 	&  	&  		& -\xi_1 	& 0 	&  -{\tilde t}_n \\
 	&  	&  		&  		& \xi_3 & 0 \\
 	&  	&  		&  		& 	& -\xi_3 \\
\end{array}
\right)
\nonumber
\eea
where one easily completes the matrix knowing that $M^t = M^*$
and the energies and gaps written below (taking the lattice parameters
$a=b=c=1$)
\bea\label{xi}
\xi_{1}&=&-\ep _{\parallel}(\cos k_x+\cos k_y) +2\ep _d \cos k_x \cos k_y -\mu
\nonumber\\
\xi_3&=& -\ep _n(\cos k_x +\cos k_y)-\mu
\nonumber\\
{\tilde t}_{\perp}&=& -\ep_{\perp} e^{i k_z d}
\nonumber\\
{\tilde t}_n&=& -\ep _ne^{i k_z d'}
\nonumber\\
\eea
with $\ep _{\a}=\d t_{\a}+J_{\a} \chi _{\a}$ for the various energies
and $\mu$ is the chemical potential which depends on $\d$.
With these expressions one can reproduce the FS of $YBCO$ while the FS
of $BSCCO$ is well represented if one neglects the next-nearest
interactions, $\ep _d=0$, and the hopping between bilayers, $\ep _n=\d t_n=0$.
The FS of figure 2 is obtained upon
adjusting the various parameters
$t_{\a}$ and $J_{\a}$ in order to reproduce experimental data
on optimally doped $YBCO$ (i.e for $\d\sim 0.25$ ) \cite{ybco}, that is (in
eV),  $t_{\parallel}\sim 0.4$,
$t_d\sim 0.1$, ${\tilde t}_{\perp}\sim 0.05$ and
${\tilde t}_{n}\sim 0.0045$ .
(We determine these values from $t_n=\Gamma^2 (a/d')^2$,
and $t_{\perp}=\Gamma^2 (a/d)^2$, using $\Gamma=1/5, d'/a=3, d/a=1$ while
$t_d$ comes from \cite{TKK}.)
As it has been pointed out in case
of $BSCCO$ \cite{KL} the hopping term between layers is taken in the form
${\tilde t}_{\perp}\sim (\cos k_x - \cos k_y)^2$.

The gaps in $M$ are given by
\bea\label{gaps}
\D_{1}&=& \D_{\parallel} (\cos k_x-\cos k_y)
+2\D_d \sin k_x \sin k_y
\nonumber\\
\D_{12}&=&\D_{\perp} e^{i k_z d}
\nonumber\\
\eea
so that the pairing along the lattice axes is of $d_{x^2-y^2}$ type
and, along the diagonals, of $d_{xy}$ type.

One can diagonalise the Hamiltonian (\ref{hk})
which gives the quasiparticle energies
\bea\label{quasiE}
E_{1,2}&=&\pm [(\xi_{1}-{\tilde t}_{\perp})^2 + (\D_1-\D_{12})^2]^{1/2}
\nonumber\\
E_{3,4}&=&\pm  [(\xi_{1}+{\tilde t}_{\perp})^2 + (\D_1+\D_{12})^2
	+\xi_3^2 +4{\tilde t}_n^2 + S^{1/2}]^{1/2}
\nonumber\\
E_{5,6}&=&\pm  [(\xi_{1}+{\tilde t}_{\perp})^2 + (\D_1+\D_{12})^2
	+\xi_3^2 +4{\tilde t}_n^2 - S^{1/2}]^{1/2}
\eea
with
\be
S=[(\xi_{1}+{\tilde t}_{\perp})^2 + (\D_1+\D_{12})^2-\xi_3^2]^2
	+8{\tilde t}_n^2[(\xi_{1}+{\tilde t}_{\perp})^2 +
	(\D_1+\D_{12})^2+2\xi_3(\xi_3+\xi_1+{\tilde t}_{\perp})]
\ee
This yields the following free energy
\be\label{FE}
F_{\D}=-2T \sum_r \sum_{\bfk }\ln[ 2\cosh (E_r({\bf k })/2T)] +
{3N\over 8}(2{\D^2_{\parallel}\over J_{\parallel}}
+2{\D^2_d\over J_d}+{\D^2_{\perp}\over J_{\perp}} )
\ee

One then minimises (\ref{FE}) which gives the
various OP. Here we restrict ourself to the above forms, see (\ref{gaps}),
for the OP, i.e. the degrees of freedom are $\D_{\parallel}$, $
\D_d$ and $\D_{\perp}$ which are real.
 There are three critical temperatures
repectively proportional to $J_{\parallel}$, $J_d$
and $J_{\perp}$; with the values that give figure 2,
$J_{\parallel}$ dominates.
In analogy with our previous OP, see (\ref{D1}), we set
\be
\left\{
\begin{array}{l}
\D_{\parallel}=\D_0\;\cos (2\a )
\\
\D_d=\D_0\;\sin (2\a)
\end{array}
\right.
\ee
Clearly the value $\a$ (mod $\pi/2)=0$ (i.e. nodal lines at
$45^{\circ}$ with respect to the lattice axes)
 will minimise the free energy since
$J_{\parallel}>>J_d$ (this can be verified e.g. upon
expanding the free energy in the small-$\D$ limit).
The additional coupling $J_d$ creates a metastable
state in which the OP has nodal lines along
the lattice axes, separated from the lowest energy state by
an energy barrier of order $J_{\parallel}$.

{}From (\ref{quasiE}) one easily verifies that the mixing of
$\D_{1}$ and $\D_{12}$ leads to two
nodal angles obtained by setting $\D_{1}\pm \D_{12}=0$ and denoted by $\a =\pm
\a_0$.
Considering the effects of the thermal
fluctuations on the layered system studied in the previous section
one gets an OP whose orientations fluctuate from
one layer to another between the values $\a=\pm \a_0$.
This will have dramatic consequences discussed in the following.

\subsection{Density of States and conductance}


In order to compute the DOS for this system    one
must have the eigenvectors of the matrix $M$. Indeed, denoting by ${\cal U}$
the unitary matrix  which
diagonalizes $ M $, the DOS in
the s-th layer is given by
\be\label{rho}
\rho_s(E) = 2\sum_{{\bf k},l}|U_{s,l}|^2
\delta (E - E_l({\bf k}))
\ee
 where
$l$ indexes the eigenvalues and E is the energy measured relatively to the
Fermi energy. In the absence of inter-bilayer coupling the
$E_l({\bf k})$ yield "nodal"
contributions for the DOS in layers 1 and 2 and "normal"  contributions for
layer 3. A nodal contribution
is obtained whenever both the $\xi$ and the $\Delta$ terms  vanish in the
quasiparticle spectrum; in
that case the DOS varies linearly with $E$ for small $E$. If only a $\xi$
term vanishes, layer 3 which has $\Delta=0$
yields a finite contribution for vanishing $E$. Including the inter-bilayer
coupling will admix these two types of
contributions such that by summing over ${\bf k}$ in any given layer
one may obtain a linear combination of "nodal" and "normal"  DOS
\cite{klemm}. ( As noted above the nodal and normal contributions  arise from
distinct ranges of integration in ${\bf k}$ space ).

With our model it is straightforward to compute ${\cal U}$ and  we only
quote the result for the DOS in layer 1 :
\be\label{dos}
\rho_1(E)=a E + b
\ee
 where $b$ is proportional to the normal state DOS in layer 3, $\rho_N$,
\be\label{b}
b =\rho_N ({t_n w\over E_3})^2
\nonumber\ee
and
\be\nonumber
 w(\vec k_0)=2 {\xi_1 + {\tilde t}_{\perp} -
 ((\xi _1 + {\tilde t}_{\perp})^2 + (\Delta _1+
\Delta _{12})^2)^{1/2}\over\Delta _1+\Delta _{12}}
\nonumber\ee
with ${\bf k}_0$ such that $\xi_3({\bf k}_0)=0$.
The offset at the Fermi energy is also present in a DOS determined
from an s-wave OP \cite{klemm} while the linear in $E$
dependence is due to the d-wave symmetry.

In single-crystal $YBCO$, the DOS has been determined from the specific heat
\cite{KAM} which is dominated
by a $T$-term at low temperature whose prefactor yields
$\rho_1(E_F)/\rho_N \sim 1/5$. This is in close agreement
with what we get for $ ({t_n w\over E_3})^2$ using
$t_n=4.5\;10^{-3}\;eV$ and $\D_{YBCO}=15meV$ \cite{L}.
On the other hand we get for $BSCCO$ $\rho_1(E_F)/\rho_N \sim
10^{-8}$ using $t_n=4\;10^{-6}\;eV$ and $\D_{BSCCO}=20meV$.

Furthermore the DOS (\ref{dos}) yields a tunneling spectrum along
the c axis shown in figure 3 at zero temperature from
\be
G(V)\sim {d\over dV}\int_0^{eV} dE \rho_0(E)\rho_1(E-eV)
\ee
where $\rho_0$ is the DOS of a conventional superconductor.
 The predicted spectrum closely ressembles
what is found in experiments on the tunneling between YBCO
and $Pb$ \cite{L,dynes}.
Above the critical temperature of $Pb$
the finite offset at zero voltage that is measured
 is given in our analysis
by the hopping energy $t_n$ (see (\ref{b})).
Below this critical temperature, the offset
induces the divergence at $\D_{Pb}=1.38$
(see figure 3). Furthermore there is
possibly a linear regime
in these experimental data coming from the d-wave symmetry of the OP of
$YBCO$. Finaly one expects a slope change at $eV=\D_{Pb}+\D_{YBCO}$.
 (In figure 3 we use $\D_{YBCO}=20meV$ which is
lower than above since we now take the value given by conductance
measurements \cite{L,dynes}.)
On the other hand no offset seems to be observed in experiments
on $BSCCO$ \cite{conduc-bscco}.


\subsection{density of states and penetration depth in the c-direction}

For a 2D d-wave superconductor one finds that the DOS
 depends linearly on E for small E (see e.g. \cite{won}
for a complete result). Indeed,
\be
\rho(E) \propto\int\int k dk d\t\quad \delta(E-\sqrt{\xi_k^2
+\Delta_{\bf k}^2})
\ee
where $\D_{\bf k}=\D_0(T) \cos (2\t)$.
We now compute the contribution to $\rho$  near
$\t={\pi\over 4}\;(mod\;{\pi\over 2}) $.
Setting $\t={\pi\over 4}-\t '$,
$x=\xi$, $y=2\Delta_0 \t '$, $r=\sqrt{x^2+y^2}$ and $\p=\tan^{-1}(y/x)$ we
obtain
\be
\rho(E) \propto\int\int rdrd\p \quad \delta(E-r)
\ee
Thus the linear behaviour is due to the nodal structure of $\D_{\bf k}$
which lead us to address the question of the effect of the thermal
fluctuations on the DOS and consequently on the penetration depth.

As discussed in the previous section, for bilayers, the contribution at
$\t={\pi\over 4}$ $\;(mod\;{\pi\over 2})$ evolves into two
modes at $\t={\pi\over 4}\pm\alpha_0 \;(mod\;{\pi\over 2})$. At $T=0$ the
inter-bilayer hopping term $t_n$ couples the bilayers
and selects the $+\alpha_0$
or $-\alpha_0$ solution to form a coherent state. For $T>t_n$ however the
bilayers are decoupled. The system in effect behaves in a 2D fashion. As a
result, the contribution to the DOS along c coming
from a particular orientation
of the nodal line -- say $\t={\pi\over 4}-\alpha_0$ -- is given by
\be
\rho_c(E,z)  \propto\int d\p\int_{\epsilon(z)} rdr \quad \delta(E-r)
\ee
with
$\epsilon(z) = 0$ or $\Delta_0 \alpha_0$ so that
\be
\begin{array}{lll}
\rho_c(E,z)\propto  E &{\rm for}& E>\epsilon\\
\rho_c(E,z)=  0 &{\rm for}& E<\epsilon
\end{array}
\ee
The average DOS is thus given by
$\rho_c(E)={<\rho_c(E,z)>}_{\epsilon}$ where
$<..>_{\epsilon}$ denotes averaging over
the  $\epsilon$ distribution function. Since the nodal lines
may take on two
positions the system is equivalent to an Ising problem. The energy barrier
between the two equilibrium positions in a bilayer is of
the order of $\Delta_0
\alpha_0$, (and can be estimated to $\sim 40 K$ using realistic values for
$\Delta_0\sim 25 meV$ and for $\alpha_0\sim 10^{\circ}$).
The energy barrier in
the vertical direction is of the order of $t_n$.
We note that for $T\ge \Delta_0
\alpha_0$ contributions to the DOS are not dominated
by the nodal lines anymore, and
gapped parts in ${\bf k}$ space  must be included.
In that limit the Ising analogy ceases
to be valid. In the Ising regime ($T< \Delta_0\alpha_0$)
however we may write
\be
P(\epsilon(z))={1\over\sqrt{2\pi\sigma^2}} e^{{-\epsilon^2\over 2\sigma^2}}
\ee
 where
the variance $\sigma\propto {\chi\over L}$ (the susceptibility divided by the
linear size along the c axis). Thus
\be
\rho_c(E)={E\over\sqrt{2\pi\sigma^2}}
\int_0^E e^{{-\epsilon^2\over 2\sigma^2}}d\epsilon
\ee
 If the susceptibility is
large enough, as is the case for a quasi 2D system and/or if L is small enough
as is the case for thin films, one may have a situation where $E\propto T$ and
$\sigma>>T$ so that
\be\label{E2}
\rho_c(E)\propto E^2
\ee
 This situation may correspond to high
$T_c$ films and to the bismuth, mercury, and thallium compounds which behave as
fairly 2D systems. On the other hand for large $L$ and/or
small $\chi$ as is the
case for single crystals and/or more isotropic compounds such as YBCO one
recovers
\be\label{E}
\rho_c(E)\propto E
\ee
 Inasmuch as in the thermal fluctuation dominated regime the penetration
depth is given by
\be
\lambda_c^{-2}(T)=\lambda_c^{-2}(0)[1-2\int-{\partial f\over\partial
E}{\rho_c(E)\over\rho_N(0)}dE]
\ee
 where $f(E)={1\over exp(\beta E)+1}$ and
$\rho_N(0)$ is the normal state DOS at the fermi energy, one sees that thermal
fluctuations may genuinely yield
\be\label{dlc}
\Delta\lambda_c^{-2}\propto T^2
\ee
 even in the
absence of disorder(\cite{RKSL}).

Experimentally it seems indeed
that for the 2D-like materials a
$T^2$ law is obtained for the c-axis penetration depth.
For $YBCO$ the situation is
still not settled. Mao et al. recently reported findings
of a linear $T$ dependence
for a $YBCO$ single crystal
\cite{mao}
whereas the UBC group seems to find a more quadratic dependence
\cite{lambda1}.
Moreover experiments on $YBCO$ powders also see
a $T^2$ dependence
that is shown to be intrinsic \cite{guerin}.
As was pointed out in \cite{guerin}
the measured $\lambda_{\parallel}$ combines $\lambda_{ab}$
and $\lambda_c$ which is much larger.
So even if the present model would give a linear behaviour
for $\D\lambda_{ab}$, one could find $\D\lambda_{\parallel}\propto T^2$
in case of a quadratic $\D\lambda_c$.

\section{conclusion}

In this paper we consider superconductors consisting
of layers with a d-wave OP that varies in  momentum
space according to
\be\label{D3}
\Delta_{\bf k} =\Delta_0(T)\cos[2(\t-\a)]
\ee
($\tan\t={k_y\over k_x}$) where $\a$ gives the nodal lines orientation.
As for the phase of the OP, $\a$ may fluctuate in space.
We consider an $\a$ that is homogeneous in the layers but can thermally
fluctuate in the c-direction. We show how the value of $\a$ is
fixed with respect to crystallographic axes
when a reasonable band structure is taken into account ( this leads to
anisotropies of the Fermi
surface ). We use an extended t-J Hamiltonian with $CuO_2$ bilayers
separated by a charge reservoir, yielding a FS for
$YBCO$ in fair agreement with that mapped out by ARPES measurements.
 This leads to various possible pairings
of the type of (\ref{D3}) with different $\a$.
Using the key idea of this paper of combining the
inexpensive spatial fluctuations
of $\a$ with the fact that the OP can choose between
several orientations, we show that the
thermal fluctuations of $\a$ change the $T$ dependence
of $\D\lambda_c$ from linear to quadratic. Our model predicts
this quadratic behaviour for thin films and quasi-$2D$ systems
where it is usually attributed to the disorder (see \cite{scal}
for a review).
We also show how the presence of the reservoir layers is revealed by
conductance and specific heat measurements.

Here we did not consider the orthorhombicity that distinguishes $YBCO$,
from e.g. $BSCCO$, that is,
asymmetric d-wave amplitudes in the $a$ and $b$ directions.
Then one gets a mixture of $d_{x^2-y^2}$ and $s_{x^2+y^2}$
(called extended s-wave) symmetries.
If the s-wave component is small
one still has node lines. In twined samples there will
be domains of different orientation of these lines,
an issue that we leave for a later publication.

Finaly let us consider the case of a pure extended s-wave
OP sometimes invoked for $BSCCO$. It has nodes and
one can easily show that they yields a linear behaviour of the DOS
near the Fermi energy.
Since there are thermal fluctuations in the
orientation of the OP as well (see section 2), one can conclude that the above
results for the penetration depth persist.

\section*{Acknowledgments}

Fruitful discussions with Sebastian Doniach,
 Walter Hardy, Jerome Lesueur and Kathreen A. Moler
are gratefuly acknowledged.

\section*{Appendix}

In the following we give a summary of the derivation of
the LG-equations for a general OP in the vicinity of
$T_c$. As in section 2 we first derive the equations
in a layer and then consider the LD-model.
The OP is defined by
\be\label{a1}
\mitD({\bf r},{\bf r}')=T\;V({\bf r}-{\bf r}')\sum_{\o}
F^+_{\o} ({\bf r},{\bf r}')
\ee
where $V$ is the attractive two-body interaction of
 the weak-coupling theory and $\o=\pi T(2n+1)$.
 We start from the Gor'kov
equation of motion for the normal and anomalous
Green functions (in the absence of a magnetic field)
\be\label{a2}
\left\{
\begin{array}{l}
(i\o -\xi_{\bf k }) G_{\o} (\bfk ,\bfk ')+
\sum_{\bf q}\mitD (\bfk ,{\bf q}) F^+_{\o} ({\bf q},\bfk ')=
\d ^2(\bfk -\bfk ')
\\
(i\o +\xi_{\bfk }) F^+_{\o} (\bfk ,\bfk ')+
\sum_{\bf q}\mitD^*(\bfk,{\bf q}) G_{\o} ({\bf q},\bfk ')=0
\end{array}
\right.
\ee
where we omit the spin variable to simplify. The energy
is described  by $\xi(\bfk )=\epsilon(\bfk )-\mu$
with $\epsilon (\bfk ) = (k_x^2+k_y^2)/ (2 m)$.
As usual one introduces the Green function of the electron
 in the normal state
\be\label{a3}
G^0_{\o} (\bfk ,\bfk ')={{\d^2(\bfk -\bfk ')} \over {i\o -\xi_{\bf k}}}
\ee
and one expands in power of $\mitD$. One gets
\bea\label{a4}
\mitD^*(\bfk _1,\bfk _2)=T\sum_{\o}\sum_{\bfk ,\bfk '}
\d^2(\bfk _1+\bfk _2-\bfk -\bfk ')V(\bfk -\bfk _1)
\left\{
{{\mitD ^*(\bfk ,\bfk ')} \over {(i\o -\xi_{\bf k })(-i\o -\xi_{\bf k '})}}
\right.\nonumber\\ \left.
-\sum_{{\bf q},{\bf q}'}
{{\mitD ^*(\bfk ,{\bf q})\mitD ({\bf q},{\bf q}')\mitD ^*({\bf q},\bfk ')}
\over
{(-i\o -\xi_{{\bf k}})(i\o -\xi_{{\bf q}})
(-i\o -\xi_{{\bf q}'})(i\o -\xi_{{\bf k '}})}}
\right\}
\eea
We shall write $\mitD (\bfk _1,\bfk _2)=
\mitD(\bfk _1-\bfk _2,\bfk _1+\bfk_2)$ which is the
notation of section 2.
Then we make the substitution in various quantities
\be\label{a5}
\xi_{{\bf k} +{\bf q}}=\xi _{\bf k } + |v|{\bf q}.{\hat {\bf k} }
+ {{{\bf q}^2}\over {2m}}\;\;\;\;{\rm with}\;\;\;
{\hat {\bfk}}={\bfk \over {|\bfk|}}\;,\;|v|={{|\bfk |}\over m}
\ee
and we expand in powers of ${\bf q}$ (we set $D_{\bf k }=i\o -\xi_{\bf k })$
\bea\label{a6}
{ 1\over {i\o -\xi_{{\bf k }+{\bf q}}}}\simeq {1\over D_{\bf k }}
\left\{
1+{|v|\over D_{\bf k }}{\bf q}.{\hat {\bfk} } +
{{|v|^2}\over {D^2_{\bf k }}}({\bf q}.{\hat {\bfk} })^2 +
{{{\bf q}^2}\over {2mD_{\bf k }}}+
{{|v|^3}\over {D^3_{\bf k }}}({\bf q}.{\hat {\bfk} })^3 +
{|v|\over {D^2_{\bf k }}}{{\bf q}^2\over m} ({\bf q}.{\hat {\bfk} })
\right.\nonumber\\ \left.
+{{|v|^4}\over {D^4_{\bf k }}}({\bf q}.{\hat {\bfk} })^4+
{{3|v|^2}\over
{D^3_{\bf k }}}{{{\bf q}^2}\over {2m}} ({\bf q}.{\hat {\bfk} })^2 +
{{{\bf q}^4}\over {4m^2D^2_{\bf k }}}
\right\}
\eea
In the following we shall keep only the dominant terms without
loss of generality. This becomes valid when $\o^2<<\mu^2$
which leads to substitute factors $v_F=(\mu /2m)^{1/2}$ for
$|v|$. Furthermore one substitutes the integration
$m\int d\xi _{\bf k }\int d\t$ for $\sum_{\bf k }$ to which
the terms with an even power of ${\hat {\bfk} }=(\cos \t,\sin \t)$
won't contribute. Thus we only keep in (\ref{a6}) three terms,
the order 0 in ${\bf q}$ and the first term written above at the
order 2 and at the order 4.

Then the particular forms of the OP and the pairing potential, see
(\ref{D1}), (\ref{D2}) and below (\ref{psi}), are to be introduced.
Upon taking the Fourier
transform with respect to the center-of-mass coordinate
${\bf q}$, which gives gradients $\naR$ of $\mitD$,
one gets (\ref{eqgap}).

In case of layered systems the gap equations can be
derived as above,
with the energy (\ref{eps}). In the expansion
of (\ref{a3}) we keep at each order in $({\bf q},q_z)$
the dominant terms in the limits $|\o |\sim T_c<<\mu \sim v_F^2$
and $K<<T_c$ at finite $K^2/(T_cv_F^2)$, that is,
\bea\label{a7}
{1\over {i\o -\xi_{{\bf k} +{\bf q}}}}\simeq {1\over D_{\bf k }}
\left\{
1+ {{|v|^2}\over {D^2_{\bf k }}}({\bf q}.{\hat {\bfk} })^2 +
{K^2\over D_{\bf k}} [1-\cos (dq_z)]
\right.\nonumber\\ \left.
+ {{|v|^4}\over {D^4_{\bf k }}}({\bf q}.{\hat {\bfk} })^4 +
{{6v_F^2K^2}\over D^4_{\bf k }}({\bf q}.{\hat {\bfk} })^2 [1-\cos (dq_z)]
\right.\nonumber\\ \left.
+ {{3K^4}\over {16D^2_{\bf k }}}[{15\over 2}-13\cos (dq_z)+
8 \cos (2dq_z) -3\cos (3dq_z) +{1\over 2}\cos (4dq_z)]
\right\}
\eea
As before we also anticipated upon supressing terms that
disapear after an integration over $(|\bfk |{\hat {\bfk}},k_z)$. One
substitutes (\ref{a7}) in (\ref{a4}) to get the gap equations
and finaly the free energy densities (\ref{fn}) and (\ref{fn1}).

\vspace{28cm}
\section*{Figure caption}

\vspace{0.5cm}
{\bf Figure 1 :}
\vspace{0.5cm}

The hopping and exchange energies on a bilayer
structure (1-2) separated by charge reservoir (3).
The summation in the Hamiltonian (15) is restricted
to nearest and next nearest neighbors witin a layer
(right figure) and to nearest neighbors in adjacent layers
(left figure).

\vspace{0.5cm}
{\bf Figure 2 :}
\vspace{0.5cm}

The Fermi surface of $YBCO_7$ for a single bilayer in the $\Gamma
X Y $ quadrant. The
solid  (resp. dashed) line corresponds to a dispersion relation of the form
$\xi_1$+ (resp. - ) \~t$_{\perp}$ (see eqn (18)). The degeneracy of the two
branches at ${\Pi\over 2},{\Pi\over 2}$ is lifted by \~t$_n$.

\vspace{0.5cm}
{\bf Figure 3 :}
\vspace{0.5cm}

Zero temperature tunneling conductance along the $c$ direction for
a $YBCO/Pb$ junction. The DOS for $YBCO$ near the Fermi energy is given by eqn
(21). The conductance diverges at the $Pb$ gap taken to be 1.38 meV.

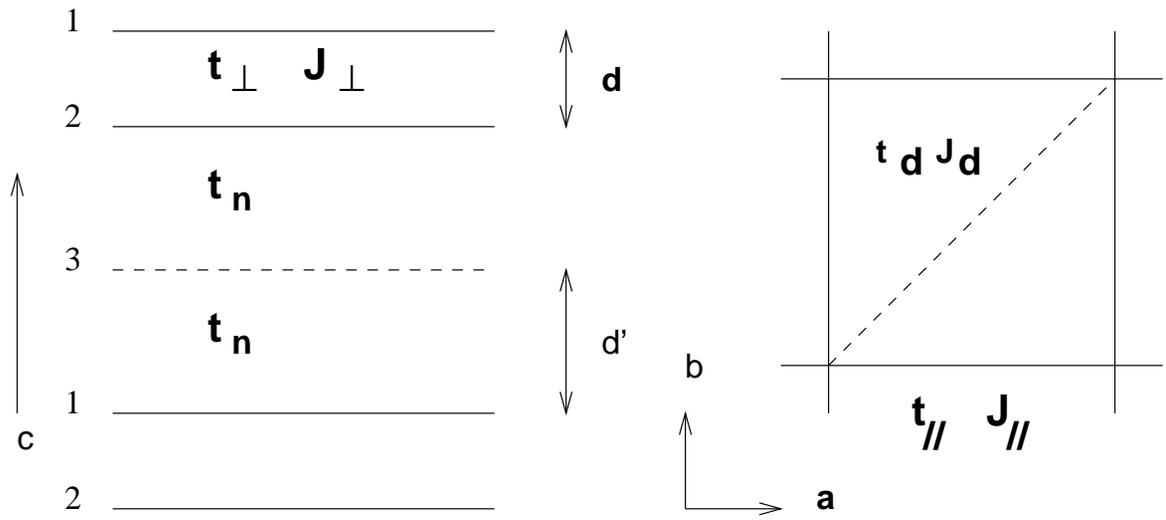
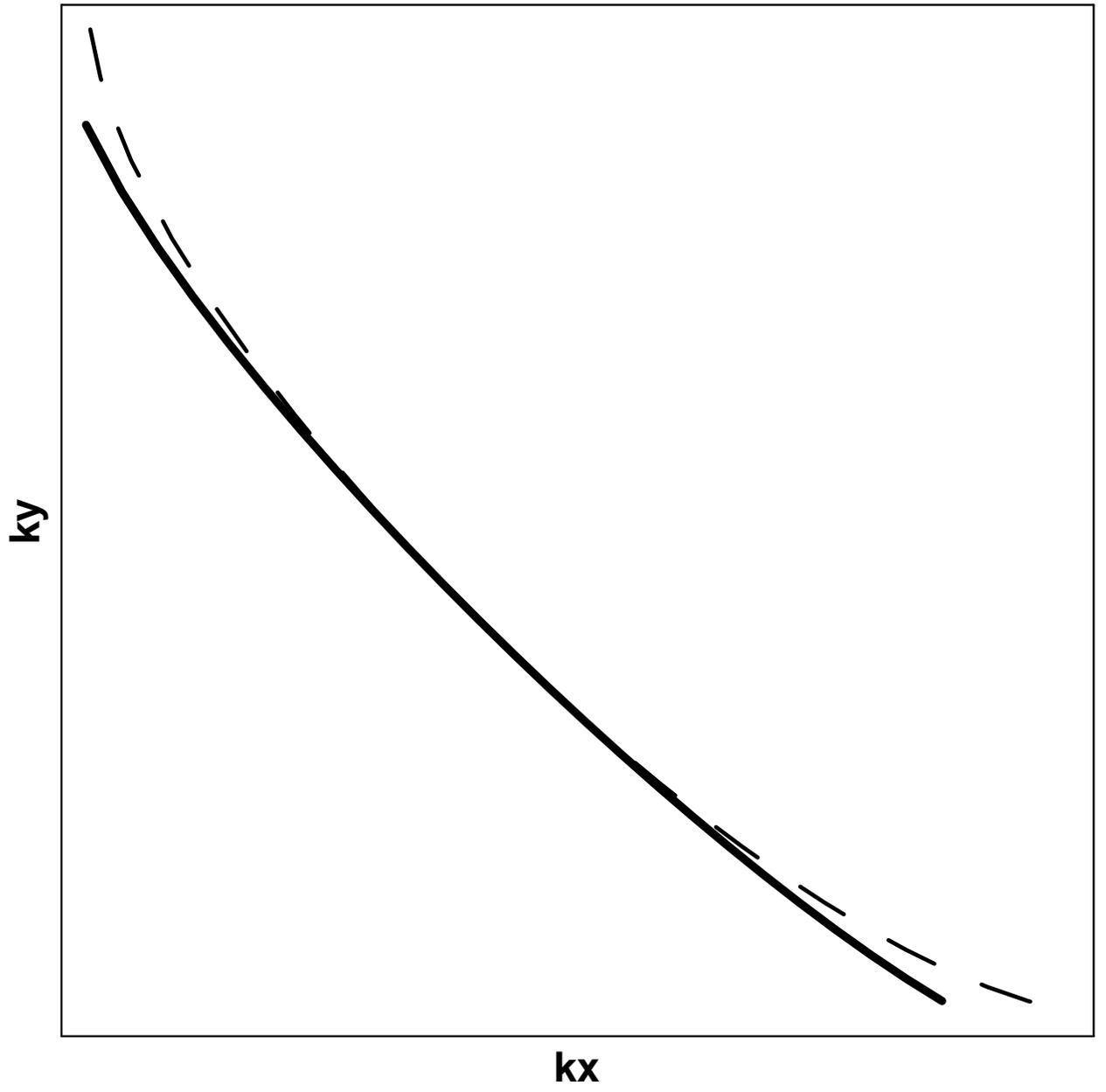
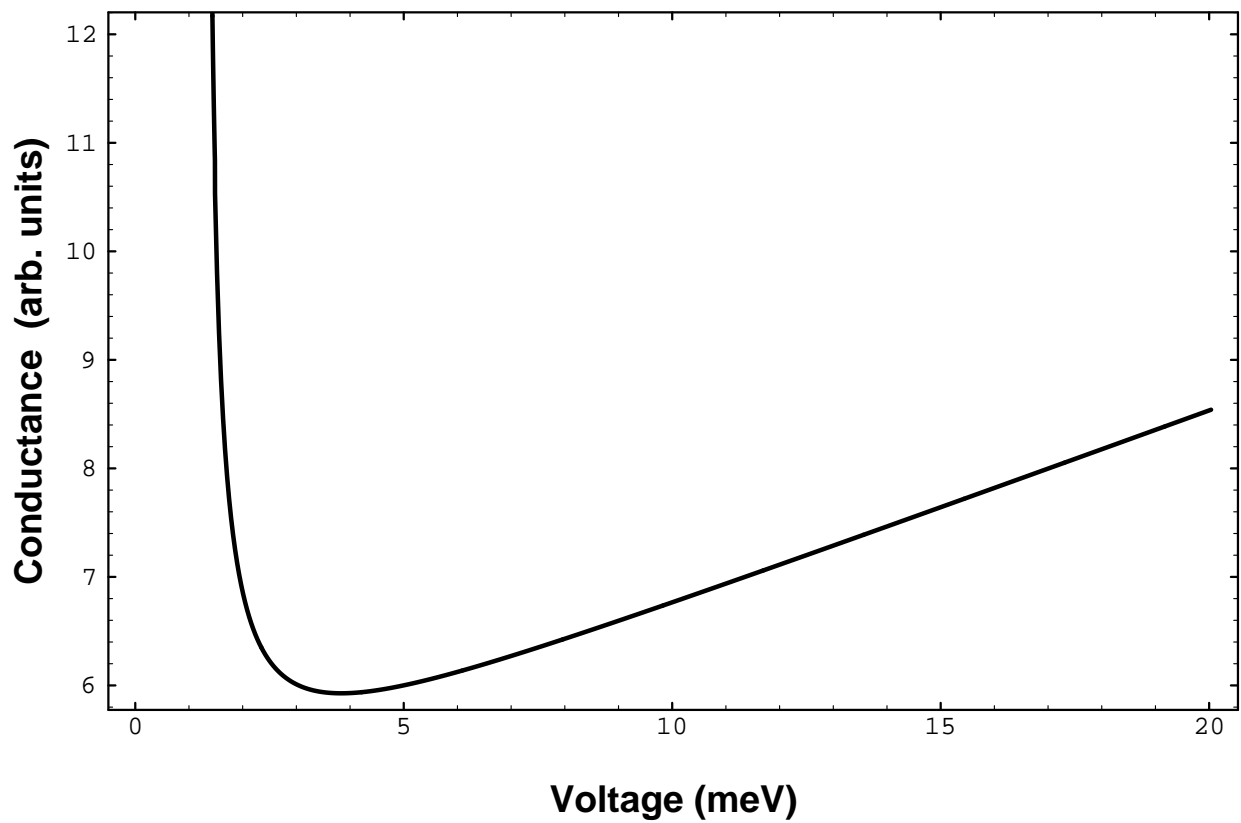
\end{document}